\documentclass[aps,preprint,showpacs,pra]{revtex4}
\usepackage{epsfig,amssymb,amscd}
\begin{document}
\bibliographystyle{prsty}

\title{\bf Readout scheme of the fullerene-based quantum computer by a single electron transistor}
\author{M. Feng $^{1,2}$ and J. Twamley $^{1}$} 
\affiliation{$^{1}$ Department of Mathematical Physics, National University of Ireland, Maynooth, Co. Kildare, Ireland \\
$^{2}$ Wuhan Institute of Physics and Mathematics, Chinese Academy of Sciences, Wuhan, 430071, China}
\date{\today}

\begin{abstract}

We propose a potential scheme, based on the achieved technique of single electron transistor (SET), to 
implement the readout of electronic spin state inside a doped $C_{60}$ fullerene by means of the magnetic dipole-dipole coupling and  
spin filters. In the presence of an external magnetic field, we show how to perform the spin state detection by 
transforming the information contained in the spin state into the tunneling current. The robustness of our scheme against sources 
of error is discussed.
\end{abstract}
\vskip 0.1cm
\maketitle

Spin based solid - state quantum computing has drawn much attention recently due to the possibility of building  
large-scale quantum processors \cite{see}. Besides the challenge of elaborately manipulating individual qubits - encoded into single 
spin states - within small regions, a serious obstacle is to read out the information contained in each of these qubits. This readout 
is both necessary to measure the state of the completed computation and required to perform quantum error correction.

To our knowledge, there have been some proposals for single spin detection. The conversion of single spin states into charge states is 
the main idea in this respect \cite {kane,loss}. Our scheme is also based on this idea. Experimentally 
Scanning- Tunneling- Microscope electron spin resonance (ESR), has demonstrated single molecule ESR spectroscopy of iron impurities 
in Silicon \cite {rao}. However theoretical work remains to clarify for the best description of the effects \cite {balatsky} although 
experimental work
in this respect is still in rapid progress \cite{durkan}. Micro-Squids is a technique capable of distinguishing large spin difference 
($\Delta m_S \sim 30$) \cite {pakes}. Magnetic Resonance Force Microscopy, which utilizes a cantilever oscillation driven by 
spins oscillating in resonance spin, is hopeful to be an efficient single spin detector \cite {berman}, although it has not yet
reached single spin sensitivity. Recently, a single $C_{60}$ molecule 
has been used as a single electron transistor (SET) through electro-mechanical coupling \cite {park}. It may be possible to use this 
quantum electromechanical system as a single spin detector in the presence of an external magnetic field.

Our work is focused on the detection of the electronic spin states of the dopant atom inside a fullerene $C_{60}$, an essential 
ingredient in the performance 
of quantum computing based on the endohedral fullerenes $N@C_{60}$ or $P@C_{60}$ \cite {harneit,suter,jason}. 
In many of these fullerene-based designs, the qubits are encoded in nuclear spins. 
The quantum gating, based on coupling the nuclear spins via the Hyperfine interaction to the electrons and then 
coupling the electrons via a magnetic dipole interaction, is done by using NMR (i.e., nuclear magnetic resonance) and 
ESR pulse sequences. It has been checked in detail in \cite {jason}, that this fullerene - based quantum computing meets 
the requirement for a quantum computer, except the lack of an effective readout technique. Since the nuclear spin is less 
sensitive to the external environment, our detection of single spin state would be carried out on the electronic spin state. 
As the swap between the nuclear spin and the electronic spin is available, the detection of an electronic spin state is an 
efficient way to execute the readout of a fullerene- based quantum computer. 

As shown in \cite{jason},  quantum information can be encoded in electronic spin states of $|\pm 1/2\rangle$ (called 
'inner qubits') or 
$|\pm 3/2\rangle$ (called 'outer qubits'), and  quantum gating can be carried out independently with inner qubits or outer qubits. 
Therefore what we want to detect is whether the inner qubit is in  $|1/2\rangle$ or $|-1/2\rangle$, or whether the outer qubit is in  
$|3/2\rangle$ or $|-3/2\rangle$.

The endohedral fullerene systems are significantly different from the systems in \cite{kane,loss}, in that the electrons of the dopant 
atoms of $C_{60}$ cannot escape the cage while preserving their spin state. Thus many of the readout  techniques under development 
elsewhere \cite{kane,loss}, will not suit endohedral fullerenes. So the first thing we have to do is to convert the electronic spin 
state information contained within the endohedral fullerene to an outside mobile electronic spin. 
We can then detect the outside spin state by transforming the spin-based information into  current-based information via spin-filters.  

As shown in Fig. 1, our readout device is very similar to a SET, which includes source, island and drain, and the voltages between two 
of them drive the electrons to jump on or off the island. However, different from \cite {park}, in our scheme, we have two built-in spin
filters located in the source and drain respectively, which block the spin-up electrons, but only let  the spin-down electrons pass 
\cite {explain}. So we have a source which could provide down-polarized current, and the click of the detector implies the arrival of a 
down-polarized electron. Moreover, we do not consider the vibrational motion of the fullerene. We suppose that the variation of the 
vibration of the fullerene is only related to the arrival and departure, instead of the spin degrees of freedom, of the mobile electron.
The validity of this assumption will be discussed later. Furthermore, the Coulomb Blockade regime is employed \cite{boese}, which means 
the $C_{60}$ on the island cannot be charged by more than one extra electron, while an electron can jump on to the island only when the 
island is electron empty, i.e. after any electron residing on the  $C_{60}$ has jumped off.
 
{\it The key idea of our scheme is to carry out a conditional gating between the motive electronic spin (shortly called 'outside spin') 
and the stationary endohedral electronic spin  (shortly called 'inside spin')}. An essential ingredient in our scheme is the microwave 
controlled entangling of the inside spin and the outside spin. The principle is quite similar to the situation 
presented in \cite{eto}, where the the electron transport through a quantum dot entangles Hyperfine-coupled nuclear spins via a spin 
flip interaction. Since the spin information will be transferred  from the
inside spin to the outside one, we are able to get the knowledge of the inside spin from the current constituted by the mobile 
electrons. We will show that the outside spin interacts with the inside spin via a  magnetic dipole-dipole coupling, and using this we 
can carry out conditional gating between the stationary and mobile spins. 

Let us consider the singly charged island. As each of the carbon atoms of the $C_{60}$ molecule are identical, we have no idea 
which atom absorbs that 
electron. As a result, the charge can be considered to distribute uniformly on the surface of the fullerene. When an external magnetic 
field is applied, {\it the net spin effect of the jumping - on electron is equivalent to that of an electron with the same spin 
state positioned at the core of the fullerene}. Therefore, when this charged fullerene is moved close to a fullerene with 
dopant atoms, the outside spin (on the empty $C_{60}$), will interact with the inside spin (in the doped $C_{60}$), 
by magnetic dipole-dipole coupling between the geometric centers of the two fullerenes. The strength of this coupling 
is proportional to {\rm (r/nm)}$^{-3}$, and is 50 {\rm MHz} in the case of 1 {\rm nm} of inter-spin distance \cite {harneit}. 

Consider the system with these two coupled fullerenes, the Hamiltonian, in units of $\hbar=1$, is  
\begin{equation}
H_{0} = g\mu_{B}B_{1}\,\sigma^{z}_{1}\otimes I_{2} + g\mu_{B}B_{2}\,I_{1}\otimes\sigma^{z}_{2}
+ J \sigma^{z}_{1}\otimes\sigma^{z}_{2}.
\end{equation}
where the subscripts 1 and 2 correspond to inside and outside spins, respectively. $J$ is the dipole-dipole coupling strength 
between the two spins. We suppose that $\nu_{1}=g\mu_{B}B_{1}/2$, and $\nu_{2}=g\mu_{B}B_{2}/2$.
As in \cite {suter,jason}, we suppose $J < |\nu_{2}-\nu_{1}|$ (weak coupling limit) and thereby we can neglect the 
terms non-commuting with the Zeeman terms. We have the 
eigenenergies $\epsilon^1_\pm=\pm 3\nu_{1} + (\nu_{2}\pm 3J/4)$, $\epsilon^2_\pm=\pm \nu_{1} + (\nu_{2}\pm J/4)$, 
$\epsilon^3_\pm=\pm \nu_{1} - (\nu_{2}\pm J/4)$, and $\epsilon^4_\pm=\pm 3\nu_{1} - (\nu_{2} \pm 3J/4)$, 
within the Hilbert space spanned by $| \pm3/2,\pm 1/2\rangle$,  and $|\pm 1/2,\pm1/2\rangle$.

To read out the state of the inside spin by means of the current made of mobile electrons we have to carry out
two-qubit gating by ESR between the inside and outside spins. Since the outside spin is very sensitive to decoherence, 
our gating time should be shorter than the decoherence time of the outside spin. As the outside spins are well polarized and their 
interaction with the inside spin is controlled the outside spins do not constitute a source of decoherence. Potential sources of 
decoherence would be magnetic dipole interactions with unknown spin centers, both electronic and nuclear etc. 
As in \cite{jason}, we consider  the ideal case when $T_2\sim T_1$ for the electrons (in the case of zero spin density). 
Due to the cage effect of the fullerene, $T^{in}_{1}$ of the inside spin is longer than 1 sec \cite {knorr}, whereas $T^{out}_{1}$ of 
the outside spin may be much shorter. Anyway $T^{out}_{1}$ is much longer than $T^{out}_{2}$, which is generally 
$T^{out}_{1}\sim$ 100 $T^{out}_{2}$. We suppose that $T^{out}_{2}$ is long enough to avoid the related homogeous broadening affecting 
our scheme.  On the other hand, to implement ESR with microwave pulses, we have to avoid 
degenerate transitions. To this end, we introduce a magnetic field gradient with
$\partial B/\partial z \ge 4\times 10^{6}$ T/m,  which separates the resonance frequencies 
of $|\pm 1/2\rangle$ by $\delta=2(\nu_{2}-\nu_{1}) \ge 127$ MHz. In Table I, we list the ESR transition frequencies. 
We can classify those transition frequencies which have (have no)  $\delta$ dependence as those which
leave the inside spin unaltered (altered). This observation will play a crucial role in the following. 

Suppose that $\nu_{1}$ and $\nu_{2}$ are of the order of {\rm GHz}, and the system under consideration is in the
Coulomb Blockade regime. We define the dwell times $t$ of the mobile electron to be the duration of each electron residing on to 
the island, and suppose we can tune $t \sim 150$ ns. As part of any previous quantum computation, we will know whether the qubit 
has been encoded into the inner or outer endohedral 
electron qubit levels, i.e. $|\pm 1/2 \rangle$ or $|\pm 3/2 \rangle$. Based on this knowledge, our detection scheme can be described 
as follows: For the case of outer qubits, we send one ESR pulse every 150 $ns$ with the frequency of $2\nu_{1}+2\delta+3J/2$ for a 
certain period, say, $10^{-2}$ {\rm sec}. If no current is found behind the spin filter in the drain, the inside spin is in the
$|3/2\rangle$ state, otherwise, it is in the $|-3/2\rangle$ state. Similarly, for the inner qubits case, we send one ESR pulse every 150 
{\rm ns} with the frequency of $2\nu_{1}+2\delta+ J/2$. If after irradiation for a short period there is no current found  
behind the spin filter in the drain, the inside spin must be in the $|1/2\rangle$ state, otherwise, it is in the $|-1/2\rangle$ state.

The operation of our scheme relies on four quantities. First of all, we need 
to know the strength of the magnetic dipole-dipole coupling $J$, which determines the frequency of the ESR pulse. 
This can be determined through interrogative ESR pulse operations. Secondly, the dwell time of the mobile  
electrons on the island should be controlled so that we can have good quality two-spin gating by ESR pulses under the 
Coulomb Blockade regime. Thirdly, the linewidth of ESR pulses should  be narrower than $J$. Fourthly, the spin filters located 
in the source and drain must work well to ensure that only down-polarized electrons are transmitted.   

However, the tunneling of the mobile electrons is intrinsically stochastic, which can be described by normal distribution. 
Experimentally 
by adjusting the voltages and the electron density of the source, we can have a mean dwell time $t_{0}$. For simplicity we define 
$\alpha=\sigma/t_{0}$ where $\sigma$ is the deviation of the normal distribution. Considering this deviation, when we perform a 
spin flip of a down-polarized mobile electron, the time evolution would instead yield 
$ -i\cos(\alpha\pi/2) |\uparrow\rangle \mp\sin(\alpha\pi/2) |\downarrow\rangle$ corresponding to dwell time 
$t_{0}(1\pm\alpha)$ respectively. 

Let us estimate the implementation of our scheme under the influence of noise sources. Based on the fact that the inside spin is 
much more stable than the outside spins, we consider following Hamiltonian 
$H=\tilde{H}_{0}+H_{1}+H_{2}$, with 
\begin{equation}
\tilde{H}_{0}= H_{0} + H^{s}_{b} + H^{s}_{d},
\end{equation}
where 
$ H_{1}= \Omega_{e} (e^{-i\omega_{e}t}\sigma^{+}_{2} + e^{i\omega_{e}t}\sigma^{-}_{2})$ and 
$H_{2}= (\Gamma_{s} \sigma^{+}_{2} + \Gamma_{s}^{+}\sigma^{-}_{2}) + \Gamma_{ps}\sigma^{z}_{2}$,
with $\sigma^{k}_{2}$ $(k=z,+,-)$ being Pauli operators for the outside spin. $s=l,r$ correspond to source and drain respectively. 
$H^{s}_{b}$ and $H^{s}_{d}$ are baths associated with spin flip and dephasing of the outside spins respectively. $\omega_{e}$ is 
the frequency of ESR. $\Gamma_{s}$ and $\Gamma_{ps}$ are the couplings to the outside spins which yield spin flip and dephasing 
respectively. If the pulse is resonant with one of the transition frequencies of the outside spins (in Table 1), we suppose 
$\Omega_{e} = \Omega_{0}$. For non-resonant case, we consider $\Omega_{e} = 0$ (i.e. no interaction) due to the large detuning. 
In this case, 
given a perfectly polarized current produced in the source, since the nonresonant ESR does not flip the outside spin, the outside spin 
should still remain well polarized after the mobile electron jumps off the island. As a result, we consider $\Gamma_{ps}$ to be zero. 
This kind of problems have been well studied \cite {wise}, from which we know that the high quality measurement in the detector can 
be achieved within the time period $1/|\Gamma_{r}|^{2}$.

Our interest is more focused on the case of $\Omega_{e} = \Omega_{0}$. In the readout stage with inside and outside spins being 
initially well 
polarized, except the short period of interaction under the ESR pulse, there is no entanglement between the inside and outside spins. 
Assuming a perfectly down-polarized source, since we cannot exactly control the dwell time of the mobile electrons on the 
island, the resulting imperfect ESR generated spin flip due to the time 
deviations $\sigma$, generates a superposition state which is more sensitive to decoherence. This is the main source of error. 
To gain more physical insight in this case, we will only consider the state of the mobile electron after the irradiation by the ESR 
pulses, with the initial state being \mbox{$|\psi_\pm\rangle= - i\cos(\alpha\pi/2) |\uparrow\rangle \mp\sin(\alpha\pi/2) 
|\downarrow\rangle$}. In the Coulomb Blockade regime we examine 
the subsequent decoherence of the mobile electron  following the master equation \cite {mil},
\begin{equation}
\frac {d\rho_{2}}{dt} = \frac {\gamma_{0}}{2}(2\sigma_{2}^{-}\rho_{2}\sigma^{+}_{2} - \sigma^{+}_{2}\sigma^{-}_{2}\rho_{2} - 
\rho_{2}\sigma^{+}_{2}\sigma^{-}_{2}) - \gamma_{p} [\sigma_{2}^{z}, [\sigma_{2}^{z}, \rho_{2}]],
\end{equation}
where $\rho_{2}$ is the density operator of the outside spin. $\gamma_{0}$, and $\gamma_{p}$, are decoherence rates regarding 
$T^{out}_{1}$, and $T^{out}_{2}$ respectively.  To avoid indistinguishability in the spectrum due 
to homogeous broadening regarding $T^{out}_{2}$, we suppose $1/\gamma_{p}=25$ $ns$ \cite {ki}. Straightforward calculation shows 
in Fig. 2 that the cross terms $\rho_{2\uparrow\downarrow}$ (or $\rho_{2\downarrow\uparrow}$), related to 
$\cos(\alpha\pi/2)\sin(\alpha\pi/2)$, will decohere to zero quickly. As long as $\alpha$ is small enough, the initial spin flip 
signature will remain as the population in $|\uparrow\rangle$ will still be much larger than that of 
$|\downarrow\rangle$ even after a time evolution for 1000 ns. This signature is translated to a current signature - 
a macroscopic quantity, and the small error occurring in  individual electron flips will not significantly affect our scheme as long as 
$\alpha$ is not too large. In above mentioned case (i.e., Eq. (3)), the detected current will be  somewhat weaker than the perfect case 
($\alpha=0$). By considering the contrast of the current, however, we can still recognize the correct spin signature from the brightness
of the current \cite{expll}. Therefore, our scheme is robust due to this projective strong measurement. 

Although we considered a perfect polarized current in above treatment, there most likely will be some  small leakage of undesired 
electrons out of the spin filters  \cite {pola}. In this case,  any up-polarized component mixing in the down-polarized current 
would evolve under resonant irradiation of  ESR pulses with the frequency  $2\nu_{1}+J/2$, (see Table I). 
Employing a magnetic field gradient to break the degenerate transitions (i.e. a non-zero $\delta$), makes such evolution non-resonant 
under the above scheme. In other words, with this magnetic field gradient, our scheme is robust to the non-ideal spin-polarized current.
However, strictly speaking, even for the case of very small undesired spin-polarized current, the full numerical study of Eq. (2) is 
necessary. We will investigate this point in detail elsewhere.
 
On the other hand, our scheme also works for the imperfectly spherical fullerene case, in which additional anisotropy terms, such as
$(\sigma_{i}^{z})^{2}$ and $(\sigma_{i}^{z})^{4}$ $(i=1,2)$, should be considered in Eq. (1). This can be easily checked by 
straightforward calculation that those additional terms only change the bottom six frequencies in Table 1, instead of the top four ones. 
 
So far, our discussion has been based on the assumption that the spin degrees of freedom of the mobile electron is independent of
the vibration of the fullerene. This is true in the case of a spatially constant magnetic field. In the case of a magnetic field with 
spatial gradients, however, we should consider the effects of this magnetic gradient due to its interaction with the inside and 
outside spins assuming the fullerene is bound harmonically in space. It yields a shift of position  
$\Delta z = (2g\mu_{B}/k)\partial B/\partial z$, with $k=70$ N/m, being the force constant of the binding harmonic oscillator force 
\cite {park}.  We get $\Delta z=2.1\times 10^{-7}$ pm, which is negligible compared to the distance variation $\delta=4$ pm caused by 
the Coulomb interaction upon the arrival of the extra electron on the fullerene. Therefore our assumption of non spin-vibration 
coupling is reasonable and valid. 


If we suppose $t_{0}=150$ ns, then we have to emit ESR pulses sequently with each pulse length of 140 ns. To have a flip operation on 
the outside spin, $\Omega_{0}$ should be 36 MHz. This kind of ESR pulses, with linewith much narrower than J, have been achieved 
experimentally \cite {carola}.  
 
Technically, the spin filters can be built from ferromagnetic materials \cite {ferro}, or semiconductor quantum dots 
\cite {loss}, which can block up-polarized electrons with high fidelity. Experimental demonstrations of the generation of 
spin polarized current with high quality has been achieved by  \cite {pola}. 
Moreover, the required magnetic field gradients is readily available with current technique \cite {new1}, and the gate performance 
by ESR rectangular pulses can be improved by the sophisticated composed rotation method \cite {new2}. Therefore, to 
reliably carry out our scheme, on the one hand, we need a long dwell time with the deviation as small as possible. On the other 
hand, the longer decoherence time (e.g. $T^{out}_{2} > 25 ns$) of the outside spin is highly expected. 

In conclusion, we have proposed a potential scheme for the readout of single spin state inside a fullerene. By strictly 
controlling the tunneling of the electron and implementing ESR pulses, based on a SET device, we can reliably convert a spin state 
signature into a current signature, which is readable and can be detected robustly. Since SET technology is developing rapidly, 
our scheme offers a promising way for the readout of fullerene based quantum computation.  

The authors thank Marko Burghard, Jim Greer, Wolfgang Harneit, Derek Mc Hugh and Carola Meyer for helpful discussion. The work is 
supported by EU Research Project QIPDDF-ROSES under contract number IST-2001-37150.

\vspace{20 pt}

{\bf Table I}. The resonant transition frequencies in various cases, where the first four rows are the transition frequencies 
of the outside spin with respect to a certain inside spin, while the final six rows list the transition frequencies 
of the inside spin with respect to a certain outside spin.

\vspace{20 pt}

\begin{tabular}{|l|l|l|}
\hline
Inside Spin ~~~ & Outside Spin ~~~& Transition Frequency \\
\hline
$|3/2\rangle$ & $|1/2\rangle\leftrightarrow|-1/2\rangle$ & $2\nu_{1}+2\delta + 3J/2$ \\
$|1/2\rangle$ & $|1/2\rangle\leftrightarrow|-1/2\rangle$ & $2\nu_{1}+2\delta + J/2$ \\
$|-1/2\rangle$ & $|1/2\rangle\leftrightarrow|-1/2\rangle$ & $2\nu_{1}+2\delta - J/2$ \\
$|-3/2\rangle$ & $|1/2\rangle\leftrightarrow|-1/2\rangle$ & $2\nu_{1}+2\delta - 3J/2$ \\
$|3/2\rangle\leftrightarrow|1/2\rangle$ & $|1/2\rangle$ & $2\nu_{1} + J/2$ \\
$|1/2\rangle\leftrightarrow|-1/2\rangle$ & $|1/2\rangle$ & $2\nu_{1} + J/2$ \\
$|-1/2\rangle\leftrightarrow|-3/2\rangle$ & $|1/2\rangle$ & $2\nu_{1} + J/2$ \\
$|3/2\rangle\leftrightarrow|1/2\rangle$ & $|-1/2\rangle$ & $2\nu_{1} - J/2$ \\
$|1/2\rangle\leftrightarrow|-1/2\rangle$ & $|-1/2\rangle$ & $2\nu_{1} - J/2$ \\
$|-1/2\rangle\leftrightarrow|-3/2\rangle$ & $|-1/2\rangle$ & $2\nu_{1} - J/2$ \\
\hline
\end{tabular}

\begin{figure}[p]
\begin{center}
\setlength{\unitlength}{1cm}
\begin{picture}(2,15)
\put(-4,3){\includegraphics[width=12cm]{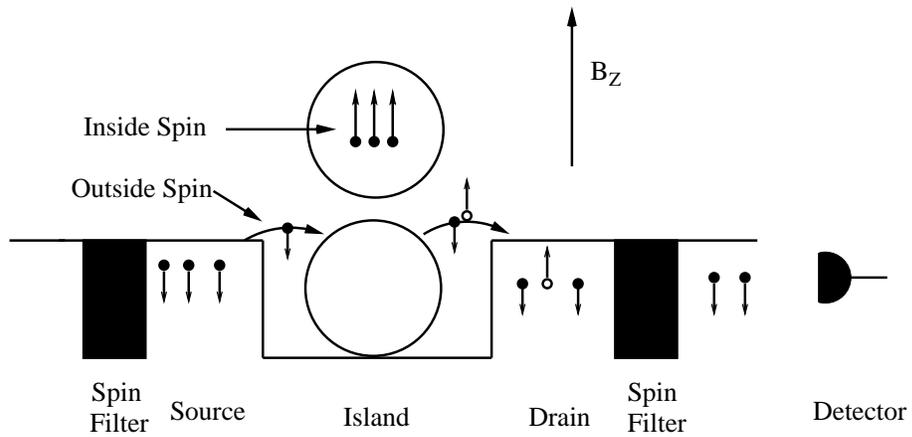}}
\end{picture}
\end{center}
\caption{Schematic diagram of our readout scheme, where the fullerene with inside spin is to be detected, and the spin-readout device
consists of source, island, drain, two built-in spin filters and a detector.  The unfilled circles denote the electrons whose
spins have been flipped by ESR pulses. The spin filter blocks spin-up electrons and only lets spin-down electrons pass. As a result, 
we have down-polarized current in the source, and the click of the detector means the presence of the filled circle electrons in 
the drain.}
\label{Fig1}
\end{figure}

\begin{figure}[p]
\begin{center}
\setlength{\unitlength}{1cm}
\begin{picture}(2,15)
\put(-4,0){\includegraphics[width=12cm]{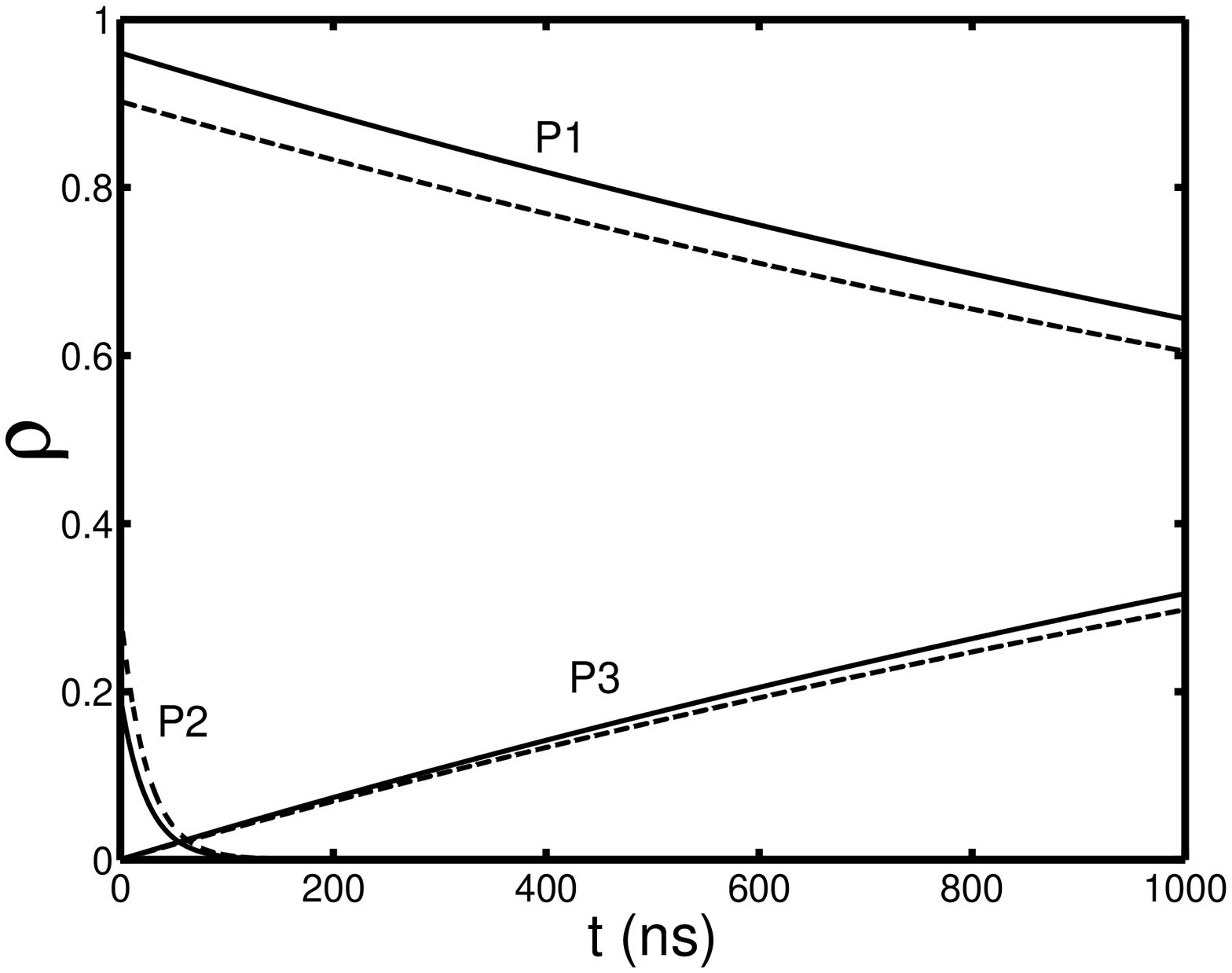}}
\end{picture}
\end{center}
\caption{Time evolution of the density matrix elements for the case of $\omega_{e}=\omega_{0}$ with the initial state 
$-i\cos(\alpha\pi/2) |\uparrow\rangle \mp \sin(\alpha\pi/2) |\downarrow\rangle$, where $1/\gamma_{p}=25$ $ns$, the solid and dashed 
curves represent $\alpha=0.1$ and $\alpha=0.2$ respectively. $P1$, $P2$ and $P3$ correspond to $\rho_{2\uparrow\uparrow}$,
$\rho_{2\uparrow\downarrow}$ (or $\rho_{2\downarrow\uparrow}$) and $\rho_{2\downarrow\downarrow}$, respectively.}
\label{Fig2}
\end{figure}

\begin{thebibliography}{22}
\bibitem{explain} There should be a way for the blocked electrons by the filter in the drain to go to ground, which is omitted 
in Fig. 1. The technical details are beyond the scope of this paper.
\end{thebibliography}
\end{document}